\documentclass[11pt, a4paper]{article}
\usepackage[utf8]{inputenc}
\usepackage[english]{babel}
\usepackage[a4paper,top=2cm,bottom=2.5cm,left=2cm,right=2cm,marginparwidth=1.75cm]{geometry}
\usepackage{amsmath,amsfonts,mathtools}
\usepackage{amssymb}
\usepackage{footmisc}
\usepackage{caption,adjustbox}
\usepackage{subcaption}
\usepackage{graphicx}
\usepackage[colorlinks=true,linkcolor=NavyBlue,citecolor=NavyBlue,urlcolor=Blue]{hyperref}
\usepackage[usenames,dvipsnames]{xcolor}
\usepackage{physics,tensor,cancel}
\usepackage{tikz}
\usetikzlibrary{calc}
\usepackage[affil-sl]{authblk}

\usepackage{comment}
\usepackage{csquotes}
\usepackage[many]{tcolorbox}    	% for COLORED BOXES (tikz and xcolor included)
\usepackage[dvipsnames]{xcolor}
\usepackage{caption,adjustbox}
\usepackage{setspace}               % for LINE SPACING
\usepackage{multicol}               % for MULTICOLUMNS

\usepackage{soul}

\usepackage{hyperref}
\usepackage[normalem]{ulem}
\usepackage{fancyhdr}
\usepackage{academicons}
\usepackage{orcidlink}
\usepackage{wrapfig,framed,caption}
\usepackage[style=numeric-comp,sorting=none,maxbibnames=9,backend=bibtex]{biblatex}
\newcommand{\ellp}{\ell_\mathrm{P}}

\newcommand{\zh}{z_H}

\newcommand{\tilxi}[1]{\widetilde{\xi}_{#1}}
\newcommand{\tiltheta}[1]{\widetilde{\theta}_{#1}}

\title{\LARGE \bf  Effective  Metric Description \\[5pt] of \\[5pt] Charged Black Holes}%Quantum Reissner–Nordström Black Holes} 

\author[1]{Mattia Damia Paciarini\thanks{{\Large \orcidlink{0009-0000-1044-341X}}
\href{mailto:damiapaciarinim@qtc.sdu.dk}{damiapaciarinim@qtc.sdu.dk}}}
\author[1,2,3]{Manuel Del Piano\thanks{{\Large \orcidlink{0000-0003-4515-8787}} \href{mailto:manuel.delpiano-ssm@unina.it}{manuel.delpiano-ssm@unina.it}}}
\author[4]{Stefan Hohenegger\thanks{{\Large \orcidlink{0000-0001-6564-0795}} \href{mailto:s.hohenegger@ipnl.in2p3.fr}{s.hohenegger@ipnl.in2p3.fr}}} 
\author[1,2,3,5]{Francesco Sannino\thanks{{\Large \orcidlink{0000-0003-2361-5326}} \href{mailto:sannino@qtc.sdu.dk}{sannino@qtc.sdu.dk}}}

\affil[1]{\mbox{\small Quantum  Theory Center ($\hbar$QTC) \& D-IAS, Southern Denmark Univ.,   Campusvej 55, 5230 Odense M, Denmark}}
\affil[2]{\mbox{\small Scuola Superiore Meridionale, Largo S. Marcellino, 10, 80138 Napoli, Italy}}
\affil[3]{\small INFN sezione di Napoli, via Cintia, 80126 Napoli, Italy}
\affil[4]{\small  Université Claude Bernard Lyon 1, CNRS/IN2P3, IP2I Lyon, UMR 5822, Villeurbanne, F-69100, France} 
\affil[5]{\small Dept. of Physics E. Pancini, Università di Napoli Federico II, via Cintia, 80126 Napoli, Italy}
\date{}

\begin{document}
\maketitle

\begin{abstract}
    \noindent Charged black holes arise as solutions of General Relativity (GR) coupled to Maxwell theory. As functions of the mass and charge, they can exhibit extremal behavior, in which case they are stable against thermal decay. (Quantum) corrections to GR are expected to alter the classical features of these objects, especially near extremality. To capture such effects in a model-independent way, we extend the Effective Metric Description (EMD) previously introduced in  \cite{DelPiano:2023fiw,DelPiano:2024gvw} for spherically symmetric and static black holes. The EMD parametrizes deformations of the metric in terms of physical quantities, such as the radial spatial distance to the event horizon. While the latter is still viable for non-extremal charged black holes, we argue that the proper time of a free-falling observer is better suited in the extremal case: we derive the necessary conditions  for the parameters of such an EMD for constructing a consistent space-time in the vicinity of the (extremal) horizon. Finally, we illustrate our framework through a concrete example, and mention implications of the Weak Gravity Conjecture on the effective metric parameters. 
\end{abstract}

\newpage

%\tableofcontents

\newpage

\section{Introduction}
In General Relativity, Reissner-Nordström black holes \cite{Reissner,Weyl,Nordstrom,Baker} are solutions of Einstein's equations coupled with Maxwell's equations. These solutions play a crucial role in advancing our understanding of black hole physics. A distinctive feature of these black holes is the presence of an upper limit on their charge, which is determined by the mass of the gravitational source. When this limit is respected, the spacetime exhibits two distinct horizons. If the charge exceeds this limit, the black hole no longer hides the singularity behind an event horizon, resulting in a naked singularity \cite{Penrose:1969pc,PhysRevLett.66.994}. When the charge reaches its maximum value, the two horizons merge, causing the surface gravity to vanish; this is known as an extremal black hole \cite{Carroll:2009maa}. 

Experimentally, strong magnetic fields have been observed around black holes, \emph{e.g.} by the Event Horizon Telescope (EHT) collaboration \cite{EventHorizonTelescope:2021bee, EventHorizonTelescope:2021srq, EventHorizonTelescope:2022urf}, further stressing the need to investigate electromagnetic interactions in the presence of black holes. 
However, at small distances, the interaction between gravitational and gauge fields remains an open question. In the absence of a fully developed theory of quantum gravity, recent efforts have shifted towards model-independent properties. Such considerations also fall in line with the Swampland program, see \emph{e.g.}~\cite{Palti:2019pca,Vafa:2005ui,vanBeest:2021lhn} for recent reviews.

It is therefore important to study possible deformations arising from modifications of Einstein gravity due to quantum effects, or other deviations from GR. Several approaches are currently under investigation  \cite{Mazharimousavi:2012ca_FofR, Campos_Delgado_2022, Donoghue:2001qc_RN_Newman, Ishibashi:2021kmf_ImprovedRN}, mostly relying on model dependent frameworks. Here, we employ and generalize the framework developed in \cite{Binetti:2022xdi,DelPiano:2023fiw,DelPiano:2024gvw} to account for  (quantum) deformations of Black Holes as classical solutions of General Relativity, via effective metrics rather than relying on computations stemming from model-dependent approximate actions and metrics \cite{bardeen1968proceedings,Hayward_2006,Dymnikova:1992ux,Bjerrum-Bohr:2002fji,Calmet:2017qqa,Calmet:2021lny, Gonzalez:2015upa,Nicolini:2019irw, Battista:2023iyu,Bargueno:2016qhu}. 
The effective metric approach provides a {\it consistent} and model-independent framework to formulate (quantum) deformations of black holes in a physically meaningful manner.
Concretely, we modify the classical metric to incorporate the aforementioned deformations. In doing so, we introduce generic deformations as functions of a physical quantity, that is compatible with the symmetries of the classical geometry (\emph{i.e.} that preserves invariance under coordinate transformations). To some degree, the choice of this physical quantity is ambiguous, but some choices are more natural than others. For non-extremal static and spherically symmetric black holes, a choice used in previous work \cite{Binetti:2022xdi,DelPiano:2023fiw,DelPiano:2024gvw,DAlise:2023hls,Hohenegger:2024kbg,DelPiano:2024nrl} is the proper spatial, radial distance from the outer horizon. However, as we shall explain in this work, for extremal black holes, this invariant breaks down at the horizon, and a natural substitute is the proper time for a free-falling observer to reach the horizon. Indeed, this invariant is well-defined at the outer horizon, and allows a coherent description of the effective metric. 
A crucial feature of all such physical quantities is that they generally depend on the geometry itself, thus posing the problem of how to compute them. To this end, assuming a certain degree of regularity, we consider series expansions around the outer horizon in a self-consistent way. Regularity conditions need to be imposed in some cases, in order to avoid physical singularities (\emph{i.e.} divergences of curvature scalars or the Hawking temperature) in physically accessible regions of space-time. The constraints obtained through this procedure provide a systematic way to classify and validate models of static and spherically symmetric metrics with a charge term. More importantly, the constraints and the relations among the series coefficients we find in our effective description do not rely on the choice of the coordinates, since they are expressed in terms of geometrical invariants.

This paper is structured as follows: In Section~\ref{Sect:RNBH}, we first summarize general features of the classical Reissner-Nordström black holes and then, we introduce the effective metric description for charged black holes. In Section~\ref{Sect:EMDDistance}, we investigate the physics near the horizon with focus on physical observables such as the modified Hawking temperature and the Ricci scalar. Imposing finiteness of these physical quantities, following \cite{DelPiano:2024gvw, DelPiano:2023fiw, Hohenegger:2024kbg} (see also \cite{DAlise:2023hls,Hohenegger:2024kbg}), we derive general constraints on the deformation functions that define the EMD. We then move to investigate the regime of small charge, that allows us to disentangle the metric deformations for the uncharged black hole in the presence of electromagnetism and determine the associated Hawking temperature in this case. We further investigate, for general charge, how the metric deformations impact the classical electromagnetic field. A crucial distinction from the uncharged case is the presence, at the classical level, of an extremal scenario, in which the internal and external horizons merge. Assuming that such a scenario occurs also at the quantum level, we show in Section~\ref{Sect:Extremal} that in this case the proper free-falling time can be used to define an EMD. Finally, we discuss some applications of the framework. In particular, we analyze the Frolov space-time \cite{Frolov:2016pav} when the extremality condition occurs, and we discuss the relation between our framework and the Weak Gravity Conjecture.    

\section{Reissner-Nordström black holes}\label{Sect:RNBH}
\subsection{Classical geometry}
The classical Reissner-Nordström geometry \cite{Reissner,Weyl,Nordstrom,Baker} represents a solution to the Einstein field equations in presence of an electromagnetic field. In the region outside of the event horizon, it reads
\begin{align}
    R_{\mu \nu} - \frac{1}{2} R g_{\mu \nu} = 8 \pi G_{\text{N}} T_{\mu \nu}   \ ,&&\text{with} && T_{\mu \nu} = \frac{1}{4 \pi}\left(F_{\mu}^{\rho} F_{\nu \rho} - \frac{1}{4} g_{\mu \nu} F^{\rho\sigma} F_{\rho\sigma}\, \right)\,, \label{EFE}
\end{align}
where $R_{\mu \nu}$ and $R$ are respectively the Ricci tensor and Ricci scalar computed by means of the space-time metric $g_{\mu \nu}$. Furthermore $T_{\mu \nu}$ is the energy-momentum tensor related to the matter and radiation fields, with $F_{\mu \nu}$ the electromagnetic field strength tensor, which is still subject to the Maxwell equations in the curved background as well as the Bianchi identity:
\begin{align}
 g^{\rho \mu} \nabla_{\rho} F_{\mu \nu} = 0\,,&&\nabla_{[ \mu} F_{\nu \rho ]} = 0 \ . \label{MFE}
\end{align}
Here $\nabla_\mu$ is the covariant derivative obtained through the metric $g_{\mu \nu}$.  The Reissner-Nordström solution then takes the following form \cite{Carroll}:
\begin{equation}
    \label{RNMetric}
    \mathrm{d}s^2 = - h(r) \mathrm{d}t^2 + f(r)^{-1} \mathrm{d}r^2 + r^2 \mathrm{d} \Omega^2 \ ,
\end{equation}
where
\begin{align}        
&h(r) = f(r) = 1 - \frac{r_\mathrm{S}}{r} + \frac{r_Q^2}{r^2} \ , &&\text{with} && \begin{array}{l}r_\mathrm{S} = 2G_{\text{N}}M\,, \\[4pt]
 r_{Q}^2 =  G_{\text{N}} Q^2 = G_{\text{N}} (q^2 + m^2) \ .\end{array} \label{ClassicalLimit}
\end{align}
Here $M$ is the mass of the black hole and $q$ and $m$ its electric and magnetic charges respectively, while $G_N$ is the Newton constant. In this paper we shall choose units, such that $G_N= \ellp^2$, where $M_{P} = \ellp^{-1}$ are respectively the Planck mass and length. In such units, the electric- and magnetic charge $q$ and $m$ are dimensionless and, following the notation in \cite{DelPiano:2023fiw}, we introduce the following further dimensionless quantities
\begin{align}
   &z \coloneqq \frac{r}{\ellp} , &&\chi \coloneqq \frac{M}{M_\mathrm{P}} \ .
\end{align}
 Consequently, equation \eqref{ClassicalLimit} can be written as follows~\cite{DelPiano:2023fiw}:
\begin{equation}
    \label{DeltaRedefinition}
    h(z) = f(z) = 1 - \frac{2\chi}{z} + \frac{Q^2}{z^2} \ .
\end{equation}
We also remark that the limit $Q \rightarrow 0$, leads to the Schwarzschild space-time, which is a solution of the first equation of \eqref{EFE} with $T_{\mu \nu} = 0$ (and a $\delta$-function singularity at the origin).

\subsection{EMD and Metric Deformations}
In order to go beyond the classical geometry of a charged black hole, we now introduce deformations in the form of an \emph{Effective Metric Description} \cite{DelPiano:2023fiw,DelPiano:2024gvw}: we keep the general form (\ref{RNMetric}) of the metric, but introduce deformations in the functions $f(z)$ and $h(z)$ in (\ref{DeltaRedefinition}). In order for these deformations to not only be compatible with the symmetries of the classical RN space-time, but also to be formulated in a universal way, we write them as functions of a physical quantity: indeed, since such a formulation is based on a measurable observable (which needs to be calculable in any given model), this approach allows to directly compare solutions stemming from different theories beyond GR. In this sense, we consider the EMD-approach universal and a means to analyse general properties of charged black holes, beyond model specific features.

Following the original work \cite{DelPiano:2023fiw,DelPiano:2024gvw}, we start by describing an EMD based on the \emph{spatial radial distance $\rho$ to the event horizon} (which is located at $z=\zh$). Keeping in mind that already the classical RN geometry features generally more than one horizon (\emph{i.e.} a radial position $z_0$ such that $f(z_0)=0=h(z_0)$), in the following we consider $\zh$ the position of the outermost horizon and we shall restrict ourselves to describing the space-time for $z\geq \zh$ (\emph{i.e.} $\rho\geq 0$) which is physically accessible. To this end, we modify the metric functions $f(z)$ and $h(z)$ in (\ref{DeltaRedefinition}):
\begin{align}
    \label{DeformedFunctions}
    h(z) = 1-\frac{2 \chi }{z} \Psi \left(\rho\right)  + \frac{Q^2}{z^2}  X \left(\rho\right), \quad
    f(z) = 1-\frac{2 \chi }{z}  \Phi \left(\rho\right)  + \frac{Q^2}{z^2}  \Upsilon \left(\rho\right) \ ,
\end{align}
with $\Psi$, $X$, $\Phi$ and $\Upsilon$ functions that define the EMD (along with the position $\zh$) and\footnote{We remark that (\ref{DeformedFunctions}) as well as (\ref{DefinitionRadialProperDistance}) are in principle well defined also for $\rho<0$, provided that the space-time geometry in the interior of the black hole is still described by a metric. In the following, however, we shall focus exclusively on $\rho\geq0$, \emph{i.e.} the space-time region outside of the black hole.}
\begin{align}
&\rho(z) \coloneqq \int_{\zh}^{z} \frac{ \dd z' }{ \sqrt{ |f(z')|} } \ ,&&\text{for} &&z\geq \zh\,.\label{DefinitionRadialProperDistance}
\end{align}
A priori $\Psi$, $X$, $\Phi$ and $\Upsilon$ are independent of each other, however, in order to describe an asymptotically flat space-time (similar to the classical RN geometry), we require
\begin{equation}
\lim_{\rho\to\infty}\Phi(\rho) = \lim_{\rho\to\infty}X(\rho) = \lim_{\rho\to\infty}\Psi (\rho) = \lim_{\rho\to\infty}\Upsilon (\rho) = 1, \quad \forall Q\in\mathbb{R}  \ .
\end{equation}
Furthermore, in order to have a horizon at $\rho=0$, we require
\begin{align}
&\Phi(0)-\frac{Q^2}{2\chi\zh}\,X(0) = \Psi(0) - \frac{Q^2}{2 \chi z_H}\Upsilon(0)\,,&& \forall  Q \in\mathbb{R}   \ .
\end{align}
Therefore, the horizon at $\rho=0$ constitutes a Killing surface with a time-like Killing vector $(K^t)^\mu = \delta^{0\mu}$:
\begin{equation}
    \left.(K^t)^{\mu} (K^t)_{\mu}\right|_{z=z_H} = \left. g_{00}\right|_{z=z_H} = - h(\zh) = 0 \ .
\end{equation}

\section{Near horizon expansion and EMD coefficients}\label{Sect:EMDDistance}
In the remainder of this paper we shall mostly be interested in developing the EMD in a region close to (but outside of) the event horizon located at $\rho=0$. In order to render the geometry fully manifest, we shall solve (\ref{DefinitionRadialProperDistance}) in the form of a series expansion, using similar techniques first developed in \cite{DelPiano:2023fiw}. 

Before entering into the details of the series expansion, we shall first simplify the metric deformations in (\ref{DeformedFunctions}). Indeed, locally, the deformation functions $\Psi$, $X$, $\Phi$ and $\Upsilon$ can  be reabsorbed into only two, which will make computations more compact in the following. Indeed, we shall consider the metric functions
\begin{align}
    f(z) = 1 - \left( \frac{2 \chi }{z} - \frac{Q^2}{z^2} \right) \Gamma \left( \rho \right), \quad
    h(z) = 1- \left(\frac{2 \chi }{z}  - \frac{Q^2}{z^2} \right)  \Omega \left(\rho \right) \ ,\label{MetricDefGen}
\end{align}
where
\begin{align}
    \Gamma \left( \rho \right) = \frac{ 2 \chi z(\rho) \Phi \left( \rho \right)  - Q^2  \Upsilon \left( \rho \right)}{2 \chi z (\rho) - Q^2},  \quad
    \Omega \left( \rho \right) = \frac{ 2 \chi z(\rho) \Psi \left( \rho \right)  - Q^2  X \left( \rho \right)}{2 \chi z(\rho) - Q^2} \ .
\end{align}
Here we have used that locally (\emph{i.e.} close to the horizon), $z$ can be written as an invertible function of $\rho$.  Furthermore, for sufficiently small values of $\rho$, we assume that $\Gamma$ and $\Omega$ can be expanded in (convergent) power series
\begin{align}
\label{SeriesExpDefFunc}
    &\Gamma \left(\rho \right) = \sum_{n = 0} \xi_n \rho^n \,,&&\text{and}&& \Omega \left(\rho \right) = \sum_{n = 0} \theta_n \rho^n\,.
    %   \xi_n= \frac{1}{n!} \left. \dv[n]{\Gamma}{\rho} \right|_{\rho=0} ,
\end{align}
The condition $f(\zh)=0=h(\zh)$ implies
\begin{equation}\label{GammaHorizonCondition}
\xi_0=\Gamma(0)=\frac{\zh^2}{2\chi \zh - Q^2}=\Omega(0)=\theta_0 \ ,
\end{equation}
such that locally, the EMD is \emph{defined} by the expansion coefficients $\{\xi_n,\theta_n\}_{n\in\mathbb{N}}$ (along with $\zh$), which we consider as physical input parameters that explicitly determine the near horizon geometry. In order to demonstrate this, we shall solve (\ref{DefinitionRadialProperDistance}) in a self-consistent fashion. However, to this end, we shall require the BH to be non-extremal, \emph{i.e.} we shall assume that the metric functions $f$ and $h$ have a simple zero at $z=\zh$ and thus admit series expansions of the form
\begin{align}
\label{DeformedNearH}
 &f(z) = f^{(1)}_H (z-z_H) + \order{(z-z_H)^{3/2}} \,,&&\text{and} &&h(z)=h^{(1)}_H (z-z_H) + \order{(z-z_H)^{3/2}} \,,
\end{align}
with $f_H^{(1)}\neq 0\neq h_H^{(1)}$ non-vanishing series coefficients.\footnote{We shall discuss in Section~\ref{Sect:Extremal} how to generalise this approach to the extremal case.} We also remark, that (\ref{DeformedNearH}) guarantees (\ref{DefinitionRadialProperDistance}) to be integrable at $\zh$ and therefore $\rho$ to be well-defined:
\begin{align}
\rho(z) = \frac{2 \sqrt{z-z_H}}{\sqrt{f^{(1)}_H}} + \order{(z-z_H)} \ .\label{dExpansion}
\end{align}
As mentioned before, locally this result can be inverted $ z(\rho) = z_H + \frac{f^{(1)}_H}{4} \rho^2 +  \mathcal{O}(\rho^3)$ and we shall (formally) write the following infinite series
\begin{equation}
\label{zExpansionInRho}
    z = z_H + \sum_{n=1}^{\infty} a_n \rho^n \ .
\end{equation}
In order to find the coefficients $\{a_n\}_{n\in\mathbb{N}}$, we need to solve (\ref{DefinitionRadialProperDistance}) order by order in $\rho$. To this end, we prefer to re-write this equation as the following differential equation 
%\begin{equation}
%    \dv{z}{\rho} = \left[ 1- \left(\frac{2 \chi}{z} - \frac{Q^2 }{z^2}\right) \Gamma (d_H + \rho )  \right]^{-1/2} \ ,
%\end{equation}
%which turns into:
\begin{equation}
\label{NonExtremeDiffEq}
     z \left( 1- \left( \dv{z}{\rho} \right)^2 \right)  =  \left(2 \chi - \frac{Q^2 }{z}\right) \Gamma (\rho ) \ .
\end{equation}
Substituting the series \eqref{SeriesExpDefFunc} and \eqref{zExpansionInRho} into \eqref{NonExtremeDiffEq}, we find:
\begin{align}
\label{Self-ConsistentCoefficients}
    \sum_{p=0}^\infty \left( 2 \chi \xi_p - Q^2 \sum_{n=0}^{p} \xi_n \mathfrak{p}_{p-n} \right) \rho^p &= z_H (1- a_1^2) + \sum_{p=1}^\infty \left[ (1-a_1^2) a_p - z_H \sum_{m=1}^{p+1} (p-m+2) m a_m a_{p-m+2} \right. \nonumber \\
    &\left. - \sum_{n=1}^{p-1} \sum_{m=1}^{n+1} (n-m+2) m a_{p-n} a_m a_{n-m+2}\right] \rho^p \ ,
\end{align}
where we have used
\begin{align}
\frac{1}{z(\rho)} = \sum_{n=0}^\infty \mathfrak{p}_n \rho^n \ ,&&\text{with}&& \begin{array}{l} \mathfrak{p}_0 = \frac{1}{z_H}, \quad \mathfrak{p}_1 = -\frac{a_1}{z_H^2}, \quad \mathfrak{p}_2 = \frac{ \left(a_1^2-a_2 z_H\right)}{z_H^3} \ , \\[20pt]
    \mathfrak{p}_p = - \frac{a_p}{z_H^2} - \frac{1}{z_H} \sum_{n=0}^{p-2} \mathfrak{p}_n a_{p-n} \ , \quad \forall p \geq 3 \ .\end{array}
\end{align}
From equation \eqref{Self-ConsistentCoefficients}, we can extract recursive conditions for the coefficients $\{a_n\}_{n\in\mathbb{N}}$: for $p=1,2$ and $3$, we find
\begin{align}
\label{FirstThreeCoeff}
    z_H^2(1-a_1^2) &= z_H 2 \chi \xi_0 - Q^2 \xi_0 \ , \nonumber \\
    -2 a_1 z_H \left(2 a_2 z_H+a_1^2-1\right) &= z_H 2\chi \xi_1 + a_1 2 \chi \xi_0 - Q^2 \xi_1 \ , \nonumber \\
    a_1^2 \left(1-10 a_2 z_H\right)-6 a_3 a_1 z_H^2+2 a_2 z_H \left(1-2 a_2 z_H\right)-a_1^4-a_2 a_1^3& = z_H 2 \chi \xi_2 + a_1  2 \chi \xi_1 + a_2 2 \chi \xi_0 - Q^2 \xi_2\ .
\end{align}
Using \eqref{GammaHorizonCondition}, the first relation indeed implies $a_1=0$, in which case the last two equations in \eqref{FirstThreeCoeff} become:
\begin{align}
\label{ConstraintOnXi1}
   & \left( 2 \chi \zh  - Q^2 \right) \xi_1 = 0 \ ,  &&\text{and} &&
    2 a_2 z_H \left(1-2 a_2 z_H\right) = \left(z_H 2 \chi - Q^2\right) \xi_2 + 2 a_2  \chi \xi_0 \ .
\end{align}
In this paper, we shall exclusively be interested in the case $Q^2\neq 2\chi\zh$,\footnote{Indeed, as we shall discuss in more detail in Section~\ref{Sect:EMDProperTime} (see in particular footnote \footref{footnoteCond}), the black holes with $Q^2\neq 2\chi\zh$ have no classical limit and therefore cannot be understood as small deformations of a geometry that appears as solution in GR. We shall not consider such scenarios in this work.} in which case the first equation requires $\xi_1=0$, which is therefore a constraint on the EMD coefficients. The second equation in (\ref{ConstraintOnXi1}) fixes the coefficient $a_2$:
\begin{align}
\label{a2Coefficient}
    a_2 =  \frac{ z_H-\xi _0 \chi +\sqrt{ \left(z_H-\xi _0 \chi \right)^2 + 4 \xi _2 z_H^2 \left(Q^2-2 \chi  z_H\right)} }{4 z_H^2} \ .
\end{align}
% Implementing the definitions of $\xi_n$ and $\zeta_n$, one finds the values of $a_2$ \manerr{redundant}
% \begin{align}
%     a_2 = \frac{z_H -\chi  \Gamma_H + \sqrt{4 z_H^2 \Gamma^{(2)}_H \left(Q^2-2 \chi  z_H \right) + \left( z_H- \chi  \Gamma _H \right)^2} }{4 z_H^2} \ . 
% \end{align}
The further coefficients of the series \eqref{zExpansionInRho} can uniquely be determined to arbitrary order using the following recursion
\begin{multline}
    a_p = \frac{1}{1-4 p \, z_H a_2} \left[ 2 \chi \xi_p -  Q^2 \sum_{n=0}^{p} \xi_n \mathfrak{p}_{p-n} + z_H \sum_{n=3}^{p-1}(p-n+2) n a_n a_{p-n+2} +\right. \\
    \left. + \sum_{n=2}^{p-2} \sum^{n}_{m=2}(n-m+2)m a_{p-n} a_m a_{n-m+2} \right] \qq{for} p \geq 3 \ .\label{Arelations}
\end{multline}

\subsection{Hawking temperature and Ricci scalar}
Besides $\xi_1=0$, we need to impose further constraints on the EMD coefficients $\{\xi_n,\theta_n\}_{n\in\mathbb{N}}$, in order to ensure regularity of the geometry: more concretely,  we require finiteness of the Hawking temperature $T_H$ and the regularity of the Ricci scalar $R=R_{\mu\nu\rho\sigma}\,g^{\mu\rho}\,g^{\nu\sigma}$ (with $R_{\mu\nu\rho\sigma}$ the Riemann tensor).

The Hawking temperature is proportional to the surface gravity at the horizon, \emph{i.e.} concretely
\begin{align}
&    T_{\text{H}} = \frac{\sqrt{f^{(1)}_H h^{(1)}_H}}{4 \pi} \,,\label{THdef}
\end{align}
with $f_H^{(1)}$ and $h_{H}^{(1)}$ given in (\ref{DeformedNearH}). In order to obtain these coefficients, we first use (\ref{zExpansionInRho}) to expand the metric functions in power series in $\rho$, supposing that the series converge in a neighborhood of the horizon
\begin{align}
\label{ExpansionsInRhof&h}
    f(\rho) &= 1 - \left(\frac{2 \chi}{z(\rho)} - \frac{Q^2}{z(\rho)^2} \right) \Gamma( \rho) = 1 + \sum_{n=0}^{\infty} \left(\sum_{k=0}^{n}    \xi_k \left( \sum_{l=0}^{n-k} Q^2 \mathfrak{p}_l \mathfrak{p}_{n-k-l} - 2 \chi \mathfrak{p}_{n-k}\right)  \right) \rho^n \ , \\
    h(\rho) &= 1 - \left(\frac{2 \chi}{z(\rho)} - \frac{Q^2}{z(\rho)^2} \right) \Omega (\rho) = 1 + \sum_{n=0}^{\infty} \left(\sum_{k=0}^{n}   \theta_k \left( \sum_{l=0}^{n-k} Q^2 \mathfrak{p}_l \mathfrak{p}_{n-k-l} - 2 \chi \mathfrak{p}_{n-k}\right)  \right) \rho^n  \ .
\end{align}
%where the first terms read explicitly
%\begin{align}
%\label{FirstTermsOfTheExpansionsInRhof}
%    f(\rho) &= \frac{\xi _2 z_H \left(Q^2-2 \chi  z_H\right)-2 a_2 \xi _0 \left(Q^2-\chi  z_H\right)}{z_H^3} \rho ^2 + \nonumber \\ & \quad + \frac{\xi _3 z_H \left(Q^2-2 \chi  z_H\right)-2 a_3 \xi _0 \left(Q^2-\chi  z_H\right)}{z_H^3} \rho ^3 + \order{\rho^4} \ ,  \\
%    \label{FirstTermsOfTheExpansionsInRhoh}
%    h(\rho) &= \frac{\theta _1 \left(Q^2-2 \chi  z_H\right)}{z_H^2} \rho + \frac{\theta _2 z_H \left(Q^2-2 \chi  z_H\right)-2 a_2 \theta _0 \left(Q^2-\chi  z_H\right)}{z_H^3} \rho ^2 +    \nonumber \\
%    & \quad + \frac{\theta _3 z_H \left(Q^2-2 \chi  z_H\right)-2 a_3 \theta _0 \left(Q^2-\chi  z_H\right)-2 a_2 \theta _1 \left(Q^2-\chi  z_H\right)}{z_H^3} \rho ^3 + \order{\rho^4} \ .
%\end{align}
%Here the $a_p$ are understood to be functions of $\{\xi_{n}\}_{n\geq 2}$ through (\ref{Arelations}), which, however, we shall not spell out in the interest of compactness of the notation. These relations can be expressed in terms of $(z-z_H)$ by means of series reversion techniques \cite{Whittaker,DelPiano:2023fiw,DelPiano:2024gvw}
%\begin{align}
%    \rho = \sum_{n=0}^{\infty} b_n (z-z_H)^{n/2} \ ,
%\end{align}
Using (\ref{zExpansionInRho}), we can then locally also write the metric functions as series expansions in $z-\zh$, for which we find to leading order
\begin{align}
\label{DeformationGeneralExpansionsZ}
    f(z) &= \frac{ z_H \xi_2 \left( Q^2  -2 \chi  z_H \right) + a_2 \left(  z_H^2 - Q^2 \xi _0 \right) }{ a_2 z_H^3} \left(z-z_H\right)+\order{(z-z_H)^{3/2}}\,,\nonumber \\
%    & \quad-\frac{\left(a_3 \xi _2-a_2 \xi _3\right) \left(Q^2-2 \chi  z_H\right)}{a_2^{5/2} z_H^2} \left(z-z_H\right)^{3/2} + \order{(z-z_H)^{2}} \ , \\
    h(z) &= \frac{\theta _1  \left(Q^2-2 \chi  z_H\right)}{\sqrt{a_2} z_H^2} \sqrt{z-z_H} + \left( \frac{\theta _2 z_H \left(Q^2-2 \chi  z_H\right)-2 a_2 \theta _0 \left(Q^2-\chi  z_H\right)}{a_2 z_H^3}+ \right.\nonumber \\
    & \quad \left. - \frac{a_3 \theta _1 \left(Q^2-2 \chi  z_H\right)}{2 a_2^2 z_H^2} \right) (z-z_H)+\order{(z-z_H)^{3/2}}\,.
    % \left(\frac{-16 a_2^3 \theta _1 \left(Q^2-\chi  z_H\right)+8 a_2^2 \theta _3 z_H \left(Q^2-2 \chi  z_H\right)}{8 a_2^{7/2} z_H^3}+ \right. \\ \nonumber
%    &  \left. + \frac{5 a_3^2 \theta _1 z_H \left(Q^2-2 \chi  z_H\right) -4 a_2 \left(a_4 \theta _1+2 a_3 \theta _2\right) z_H \left(Q^2-2 \chi  z_H\right)}{8 a_2^{7/2} z_H^3} \right) (z-z_H)^{3/2} + \order{(z-z_H)^{2}} \ .
\end{align}
Here the $a_p$ are understood to be functions of $\{\xi_{n}\}_{n\geq 2}$ through (\ref{Arelations}), which, however, we shall not spell out in the interest of compactness of the notation. For $T_{\text{H}}$ in (\ref{THdef}) to be well-defined, we first require the term of order $\order{(z-\zh)^{1/2}}$ in the expansion of $h$ to vanish. Recalling that we assumed $Q^2\neq 2\chi z_H$, this imposes the constraint $\theta_1=0$. The Hawking temperature is then well-defined and can be written as
%\begin{align}
%\label{TemperatureInGeneralExpansion}
%    T_{\text{H}} &= \frac{\sqrt{f^{(1)}_H h^{(1)}_H}}{4 \pi} \nonumber \\
%    &= \frac{\sqrt{ \left( a_2 \left(z_H^2-Q^2 \xi _0 \right) + z_H \left( Q^2 \xi _2-2 \chi \xi _2 z_H \right) \right) \left( a_2 \left( z_H^2-Q^2 \theta_0 \right) + z_H \left( \theta_2 Q^2 -\theta _2 2 \chi z_H^2 \right)\right) }}{a_2 z_H^3 4 \pi } \ .
%\end{align}
%In terms of the deformation functions, the latter can be expressed as  
\begin{align}
    T_{\text{H}}^2  = \frac{\left(z_H^2-Q^2 \xi_0 \right) \left(z_H -\chi  \xi_0 +\sqrt{4 z_H^2 \xi_2 \left(Q^2-2 \chi  z_H\right)+\left(z_H-\chi  \xi_0 \right)^2}\right)+4 z_H^3 \theta_2 \left(Q^2-2 \chi  z_H\right)}{z_H^5 16 \pi^2 } \ .
\end{align}
Since $\xi_0$ is given in \eqref{GammaHorizonCondition}, from the coefficients defining the EMD, the Hawking temperature depends only on three deformation parameters which are $z_H, \xi_2$ and $\theta_2$. Among these, we furthermore require
%From the expression of the Hawking temperature, one can extract two more constraints:
%\begin{align}
%    \fn{1}  > 0, \qq{then} T_H \in \mathbb{R} \ ,
%\end{align}
%which implies
\begin{align}
\label{ConditionsOnSecondDerivatives}
    \xi_2 \leq \frac{\left(z_H-\chi  \xi_0\right)^2}{4 z_H^2 \left(2 \chi  z_H-Q^2\right)} \ , \quad
    \theta_2 < \frac{\left(z_H -\chi  \xi_0 +\sqrt{4 z_H^2 \xi_2 \left(Q^2-2 \chi  z_H\right)+\left(z_H-\chi  \xi_0 \right)^2}\right) \left(z_H^2-Q^2 \xi_0 \right)}{4 z_H^3 \left(2 \chi  z_H-Q^2\right)} \ ,
\end{align}
to ensure that $T_H\in\mathbb{R}_+$. Finally, we also require the that the Ricci scalar $R$ is regular at the horizon. The constraints imposed above are not sufficient to make the Ricci finite. Indeed, a residual divergence is still present at the horizon, \emph{i.e.} $R \propto \frac{1}{\rho}$. The exact expression for the divergent term of $R$ is not relevant, however, it vanishes provided  
\begin{align}
    \xi_{3} = \frac{1}{2} \theta _3 \left(\frac{3 \varpi  \left(Q^2-2 \chi  z_H\right)}{ \left(\chi  z_H-Q^2\right)}-1\right)\ ,&&\text{with} &&
    \varpi = \sqrt{\frac{ \left(4 \xi _2 \left(Q^2-2 \chi  z_H\right)^3+\left(Q^2-\chi  z_H\right)^2\right)}{\left(Q^2-2 \chi  z_H\right)^2}} \ .
\end{align}
This constraint smoothly reduces to the condition found in \cite{DelPiano:2023fiw} in the limit $Q \to 0$.\\
It is useful to collect below the universal finiteness conditions on the EMD coefficients emerging in the near horizon region:
\begin{tcolorbox}[ams equation,colback=black!10!white,colframe=black!95!green]
\parbox{15cm}{${}$\\[-40pt]\begin{align}&\xi_0 = \theta_0 = \frac{z_H^2}{2 \chi z_H -Q^2}, \  \quad \xi_1 = \theta_1 = 0, \  \quad   \xi_2 \leq \frac{\left(z_H-\chi  \xi_0 \right)^2}{4 z_H^2 \left(2 \chi  z_H-Q^2\right)},\ 
\nonumber\\
& \theta_2 < \frac{\left(z_H -\chi  \xi_0 +\sqrt{4 z_H^2 \xi_2 \left(Q^2-2 \chi  z_H\right)+\left(z_H-\chi  \xi_0 \right)^2}\right) \left(z_H^2-Q^2 \xi_0 \right)}{4 z_H^3 \left(2 \chi  z_H-Q^2\right)},\,\nonumber\\
& \xi_{3} = \frac{1}{2} \theta _3 \left(\frac{3 \varpi  \left(Q^2-2 \chi  z_H\right)}{ \left(\chi  z_H-Q^2\right)}-1\right)\, , \quad\text{with} \quad \varpi = \sqrt{\frac{ \left(4 \xi _2 \left(Q^2-2 \chi  z_H\right)^3+\left(Q^2-\chi  z_H\right)^2\right)}{\left(Q^2-2 \chi  z_H\right){}^2}} 
\label{constraints}\end{align}\nonumber ${}$\\[-30pt]}
\end{tcolorbox}

\subsection{Small charge expansion}

The aim of this subsection is to provide a description of the Hawking temperature $T_{\text{H}}$ for a (quantum) deformed black hole in presence of a small charge $Q$. In this approximation, the deformations receive sub-leading corrections in $Q$: the presence of a charge enters the equations at the same order as in the classical small charge limit. We shall use a subscript $_*$ to indicate that a quantity is evaluated for $Q=0$.  In particular, the radial coordinate evaluated at the horizon in the small charge limit becomes:
\begin{align}
    z_H = z_{\ast} - \frac{Q^2}{2 \chi} + \mathcal{O} (Q^3) \ , 
\end{align}

\begin{comment}
    In the first approximation, the deformation functions can be substituted by the correspondent deformations computed at the horizon of a deformed uncharged spherically symmetric black hole:

\begin{align}
    &\Gamma_H = \Gamma_{\ast}, && \Gamma_H^{(2)} = \Gamma_{\ast}^{(2)}, 
    &\Omega_H = \Omega_{\ast}, && \Omega_H^{(2)} = \Omega_{\ast}^{(2)}.
\end{align}
\end{comment}
Implementing these approximations into the temperature $T_{\text{H}}$, one gets:
\begin{align}
    T_{\text{H}} = T_{\ast} - Q^4 \frac{ \left(8 \chi  z_{\ast} \left(\xi_{2\ast} \left(- 10 \varpi_* +32 \chi  \theta_{2\ast} z_{\ast}-6\right)-\theta_{2\ast}\right)+ 3 \varpi_* \right)}{512 \pi^2  \chi ^2 z_{\ast}^4  \varpi_* T_{\ast}  } + \mathcal{O}(Q^5) \ ,
\end{align}
in which $T_{\ast}$ represents the Hawking temperature for the uncharged black hole defined as follows:
\begin{align}
    T_{\ast} = \frac{\sqrt{ \varpi_*-8 \chi  \theta_{2\ast} z_{\ast}+1} }{4  \pi  z_{\ast}} \ , \quad \varpi_*= \sqrt{1- 32\xi_{2*} z_* \chi}
\end{align}
The previous result shows corrections to the temperature of order $Q^4$, as predicted by semi-classical approximation \cite{Balbinot:2007kr}. However, the correction differs from the classical prediction by the presence of the derivatives of the deformation functions, which enter in a non trivial way.

\subsection{Solution of the Maxwell field}
Once the divergences of the Ricci scalar have been removed, and the Hawking temperature $T_{\text{H}}$ is well-defined, one can study the behaviour of the modified electromagnetic tensor $F_{\mu \nu}$. \\
We suppose that deformations due to (quantum) gravitational effects generated by a spherical source preserve the spherical symmetry. Because of this, one may add deformations to the classical field strength tensor $F^{cl}_{\mu \nu}$ which only depends on the radial variable $z$:
\begin{equation}
\label{FST}
    F_{\mu \nu} = F^\text{cl}_{\mu \nu} + F^\text{def}_{\mu \nu} \ , 
\end{equation}
where
\begin{align}
    &E_z = F^\text{cl}_{zt} = \frac{q}{z^2}  \ , &&\text{and} && B_z = \frac{F^\text{cl}_{\theta \phi}}{z^2 \sin \theta} = \frac{m}{z^2} \ ,
\end{align}
with $Q^2 = q^2 + m^2$ as defined in \eqref{ClassicalLimit}, whereas
\begin{align}
  &E^\text{def}_z = F^\text{def}_{zt}  \coloneqq w(z) \ , &&\text{and} && B^\text{def}_z = \frac{F^\text{def}_{\theta \phi}}{z^2 \sin \theta} \coloneqq k(z) \ ,
\end{align}
in which $w(z)$ and $k(z)$ are unknown functions of the radial coordinate. The Maxwell equations can be written as:
\begin{align}
&    \partial_{\mu} \left( \sqrt{-g} F^{\mu \nu}\right) = 0  \ ,&&\text{with} &&g \coloneqq \det g_{\mu \nu} = -\frac{h(z)}{f(z)}z^4 \sin^2 \theta \ ,
\end{align}
which imply:
\begin{align}
\label{SimplifiedME}
    \left(\partial_z \sqrt{-g}\right) F^{zt} +  \sqrt{-g} \left( \partial_z F^{zt}\right) &=0\ , \\
    \label{magneticME}\left(\partial_\theta \sqrt{-g}\right) F^{\theta \phi} +  \sqrt{-g} \left( \partial_\theta F^{\theta \phi}\right) &= 0  \ , 
\end{align}
The first equation in \eqref{SimplifiedME} gives the following condition for $w(z)$:
\begin{align}
    w(z) &= -\frac{q}{z^2} + \frac{c_1}{z^2} \sqrt{\frac{h(z)}{f(z)}} 
\end{align}
where $c_1$ is a constant, which can be fixed requiring $w(z)=0$ when the deformations are switched off. This condition yields $c_1 = q$.
On the other hand, equation \eqref{magneticME} is always satisfied, independently of the choice of $k(z)$. 

Regarding the Bianchi identity
\begin{equation}
    \partial_\mu F_{\nu \rho} + \partial_\nu F_{\rho \mu} + \partial_\rho F_{\mu \nu} = 0 \ ,
\end{equation}
the only non-trivial condition is $\partial_z F_{\theta \phi} = 0$, which implies $\partial_z [k(z)z^2] = 0$ and which gives the following solution for $k(z)$
\begin{equation}
    k(z)= \frac{c_2}{z^2} \ ,
\end{equation}
where $c_2$ is an integration constant which must be determined. One can fix the latter by requiring that when $k(z) \to 0$, the classical result is reproduced. Indeed, this is satisfied if $c_2 = 0$. Then, the overall expression for the $F_{\mu \nu}$ takes the following form:

\begin{align}
    &F_{zt} = \frac{q}{z^2} \sqrt{\frac{h(z)}{f(z)}} =  \frac{q}{z^2} \frac{ \sqrt{Q^2 \Omega \left(\rho \right)-2 \chi  z \Omega \left(\rho \right)+z^2}}{\sqrt{Q^2 \Gamma \left(\rho\right)-2 \chi  z \Gamma \left(\rho\right)+z^2}} \ , &&F_{\theta \phi} = m \sin \theta \ .
\end{align}
In summary, the electric field gets modified by the factor $\sqrt{\frac{h(z)}{f(z)}}$, while the magnetic field remains unchanged respect to the classical one. This is not surprising, since the metric deformations (\ref{MetricDefGen}) where introduced in a way to keep the geometry static.

\section{Effective description of an extremal BH}\label{Sect:Extremal}
In this Section we develop an EMD for extremal (charged), spherically symmetric and static black holes. Extremality in this context amounts to assuming that the metric functions $f$ and $h$ have higher order zeros at the horizon (when seen as functions of $z-\zh$): concretely, this amounts to assuming that $ f^{(1)}_H$ in (\ref{DeformedNearH})  (and its counterpart $ h^{(1)}_H$) vanish. In more physical terms, this condition also implies that the surface gravity $\kappa\sim\sqrt{f^{(1)}_H\,h^{(1)}_H}$ of the black hole vanishes.

From a technical perspective, $ f^{(1)}_H=0$ renders the expansion (\ref{dExpansion}) (and all subsequent steps depending on it) ill-defined. This problem can also be understood from the definition of the distance in eq.~(\ref{DefinitionRadialProperDistance}): indeed, $f$ having a second order zero at $z=\zh$ introduces a non-integrable singularity at the horizon and therefore makes the distance $\rho$ of a space-time point to the horion divergent. Thus, in the case of extremal spherically symmetric and static black holes, the (spatial) distance is not a suitable physical quantity to define an EMD, at least in the immediate vicinity of the horizon. In the following, we shall therefore introduce an EMD based on the proper time of a free-falling observer (see \cite{SHinprogress} for the full EMD developed in the case of uncharged black holes), which can be generalised also to the extremal case.

\subsection{EMD based on Proper Time}\label{Sect:EMDProperTime}
\begin{figure}[h]
\begin{center}
\includegraphics[width=7.5cm]{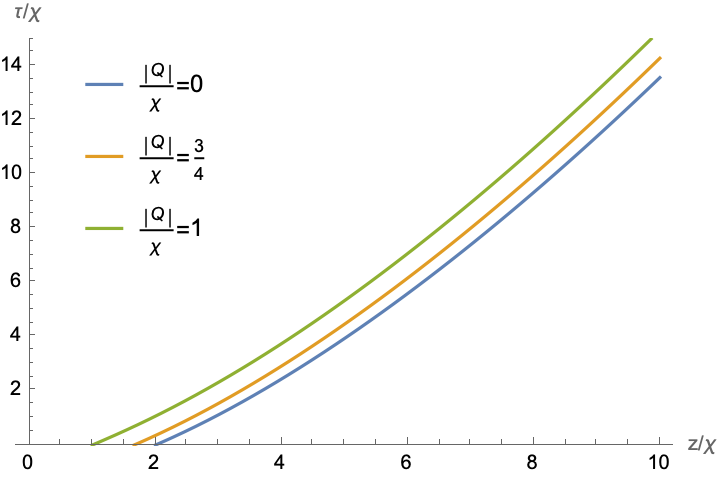} 
\end{center}
\caption{Proper time measured by a time-like observer falling along a radial geodesic from a space-time point labelled by $z$ to the horizon of a Reissner-Nordstr\"om black hole, as given in (\ref{AnalyticRN}). The case $|Q|/\chi=1$ represents the extremal case, which is still a finite function.}
\label{Fig:RNtime}
\end{figure}

We introduce the proper time $\tau$ as the time that a free-falling, time-like observer measures along a radial geodesic. Given a generic static spherically symmetric metric
\begin{align}
    \dd s^2 = - h(z) \dd t^2 + \frac{\dd z^2}{f(z)} + z^2 \dd \theta^2 + z^2 \sin^2 \theta \dd \phi^2 \ ,\label{MetricForm}
\end{align}
the infinitesimal distance for a geodesic radial motion in the equatorial plane, \emph{i.e.} for $\theta = \pi/2$, is
\begin{align}\label{defddtau}
    -\dd \tau^2 = \frac{ u \, h(z)}{f(z)} \frac{\dd z^2 }{(1+ u \, h(z))} \ ,
\end{align}
where the choice $u = + 1$ or $-1$ correspond, respectively, to either a space-like or a time-like radial geodesic. Thus, the finite proper time $\tau(z)$ that a time-like observer measures to fall from a point in space-time labelled by $z$ to the horizon at $\zh$ is
\begin{align}
\label{FreeFallingTime}
     \tau(z) = \int_{z_H}^{z} \sqrt{\frac{h(z')}{f(z')}} \frac{\dd z' }{\sqrt{1-h(z')}} \ .
\end{align}
Here the overall sign has been fixed to represent an in-falling motion along the geodesic. As an example, we consider the classical Reissner-Nordström black hole geometry, which is characterised by $f(z) = h(z) = 1 - \frac{2 \chi}{z} + \frac{ Q^2}{z^2}$. In this case, the proper time for a a radially in-falling time-like observer can be written in the form
\begin{align}
&\tau(z)=\frac{\chi}{3} \left(\mathfrak{q}^2 \left(\sqrt{2 \mathfrak{z}-\mathfrak{q}^2}-\sqrt{2 \mathfrak{z}_H-\mathfrak{q}^2}\right)+\mathfrak{z} \sqrt{2\mathfrak{z}-\mathfrak{q}^2}-\mathfrak{z}_H \sqrt{2 \mathfrak{z}_H-\mathfrak{q}^2}\right)\,,\label{AnalyticRN}%&&\text{with} && \begin{array}{l}\mathfrak{q}=|Q|/\chi\,, \\ \mathfrak{z}=z/\chi\,,\\ \mathfrak{z}_H=\frac{\zh}{\chi}=1+\sqrt{1-\mathfrak{q}^2}\,.\end{array}
   \end{align}
with the short-hand notation
\begin{align}
&\mathfrak{q}=\frac{|Q|}{\chi}\,,&&\mathfrak{z}=\frac{z}{\chi}\,,&&\mathfrak{z}_H=\frac{\zh}{\chi}=1+\sqrt{1-\mathfrak{q}^2}\,.
\end{align}   
The proper time (\ref{AnalyticRN}) is schematically shown in Figure~\ref{Fig:RNtime} for different values of the charge-to-mass ratio $\mathfrak{q}\in[0,1]$. 
The case $\mathfrak{q}=0$ corresponds to the un-charged (Schwarzschild) black hole, while the extremal case is $\mathfrak{q}=1$.\footnote{$\mathfrak{q}>1$ does not represent a black hole geometry, but rather a naked singularity, which is generally considered unphysical.} As can be seen from Figure~\ref{Fig:RNtime}, $\tau(z)$ remains a finite function (even for $\mathfrak{q}=1$) at the horizon located at $\zh=\chi\,\mathfrak{z}_H=\chi(1+\sqrt{1-\mathfrak{q}^2})$, where $\tau(\zh)=0$.
%\begin{align}
%    \tau(z) &= -\frac{1}{3 \chi ^2} \left(\chi  \left(\sqrt{\chi ^2-Q^2}+\chi \right) \sqrt{2 \chi  \left(\sqrt{\chi ^2-Q^2}+\chi \right)-Q^2} \right. + \nonumber \\
%    & \qquad\left. +Q^2 \left(\sqrt{2 \chi  \left(\sqrt{\chi ^2-Q^2}+\chi \right)-Q^2}-\sqrt{2 \chi  z-Q^2}\right)-\chi  z \sqrt{2 \chi  z-Q^2}\right) \ .
%\end{align}
%At extremality $\chi = Q$ yielding
%\begin{align}
%    \tau(z) = \frac{1}{3} \left( \sqrt{2\chi z - \chi^2} - 2 \chi \right) + \frac{z}{3\chi} \sqrt{2 \chi z - \chi^2}\ .
%\end{align}
%Having learned that the proper time is well-defined at and around the horizon in the extremal classical case, we employ it for the more general case. Preparing for the EMD near the horizon, we write 
%\begin{align}
 %   \tau_\text{tot} &= \tau_H + \tau(z)= \tau_H + \int_{z_H}^{z} \sqrt{\frac{h(z')}{f(z')}} \frac{\dd z' }{\sqrt{1-h(z')}} \ ,\label{TauDiffEq}
%\end{align}
%where $\tau_H$ is the proper time spanning from the origin to the horizon, and $\tau(z)$ is the displacement from the horizon. 
We next write the metric-deformation functions $f$ and $h$ in (\ref{MetricForm}) in terms of $\tau$ as follows
\begin{align}
    f(z)= 1- \left( \frac{2 \chi}{z} -\frac{Q^2}{z^2}\right) \widetilde{\Gamma} (\tau) \qq{and} 
    h(z)=1- \left( \frac{2 \chi}{z} -\frac{Q^2}{z^2}\right) \widetilde{\Omega} (\tau)\ .\label{fhFunctions}
\end{align}
Here, the functions $\widetilde{\Gamma} (\tau)$ and $\widetilde{\Omega} (\tau)$ (along with $\zh$) define the EMD based on the proper time $\tau$. In order to avoid confusion with the EMD based on the spatial distance to the horizon,\footnote{Since both $\rho$ and $\tau$ are monotonic functions of $z$ in the vicinity of the horizon and finite in the non-extremal case, we remark that the EMDs based on the two quantities are therefore equivalent and one can be expressed in terms of the other. As mentioned before, this is no longer possible in the extremal case.} as discussed in Section~\ref{Sect:EMDDistance}, we have renamed the deformation functions $ \widetilde{\Gamma} (\tau)$ and $ \widetilde{\Omega} (\tau)$ respectively, and we expand them as power series of the proper time from the horizon in the following manner
\begin{align}
&\widetilde{\Gamma} (\tau) = \sum_{n=0}^\infty \tilxi{n}\tau^n \ , && \widetilde{\Omega} (\tau) = \sum_{n=0}^\infty \tiltheta{n}\tau^n \ ,&&\text{with} &&\tilxi{n}\,, \tiltheta{n}\in\mathbb{R}\,.\label{zExpansiontau}
\end{align}
In order to solve (\ref{FreeFallingTime}) iteratively, we also introduce the expansions
\begin{align}
&z = z_H + \sum_{n=1}^{\infty} \widetilde{a}_n \tau^n \,,&&\tau(z)=\sum_{n=1}^\infty \widetilde{b}_n\,(z-\zh)^n\,,&&\text{with} &&\widetilde{a}_n\,,\widetilde{b}_n\in\mathbb{R}\,,\label{TauExpansion}
\end{align}
where the coefficients $(\widetilde{a}_n)_{n\in\mathbb{N}}$ are implicitly determined by (\ref{FreeFallingTime}), while $(\widetilde{b}_n)_{n\in\mathbb{N}}$ are fixed through series inversion \cite{Whittaker,DelPiano:2023fiw,DelPiano:2024gvw}. Notably for the first few instances we find (see also \cite{SHinprogress})
\begin{align}
&\widetilde{b}_1=\frac{1}{\widetilde{a}_1}\,,&&\widetilde{b}_2=-\frac{\widetilde{a}_2}{\widetilde{a}_1^3}\,,&&\widetilde{b}_3=\frac{2\widetilde{a}_2^2-\widetilde{a}_1\widetilde{a}_3}{\widetilde{a}_1^5}\,,&&\widetilde{b}_4=\frac{5\widetilde{a}_1\widetilde{a}_2\widetilde{a}_3-5\widetilde{a}_2^3-\widetilde{a}_1^2\widetilde{a}_4}{\widetilde{a}_1^7}\,.\label{SeriesInversion}
\end{align} 
In order to represent a black hole horizon, the metric functions $f$ and $h$ in (\ref{fhFunctions}) need to vanish for $z\to \zh$, which is equivalent to $\tau\to 0$. Since (\ref{zExpansiontau}) and (\ref{TauExpansion}) lead to 
\begin{align}
&f(z)=\left(1-\frac{\tilxi{0}(2\zh\chi-Q^2)}{\zh^2}\right)+\mathcal{O}(z-\zh)\,,&&h(z)=\left(1-\frac{\tiltheta{0}(2\zh\chi-Q^2)}{\zh^2}\right)+\mathcal{O}(z-\zh)\,,\label{fhzExpansion}
\end{align}
$\tilxi{0}$ and $\tiltheta{0}$ cannot be arbitrary, but need to satisfy
\begin{align}
\tilxi{0}=\frac{\zh^2}{2\zh\chi-Q^2}=\tiltheta{0}\,.\label{Condxi0}
\end{align}
Here, for simplicity, we restrict to the case $2\zh\chi\neq Q^2$.\footnote{In the classical case, this condition is equivalent to $\mathfrak{q}^2\neq 2(1+\sqrt{1-\mathfrak{q}^2})$, which is always satisfied over the reals, notably for $\mathfrak{q}\in[0,1]$. Thus, deformed black holes with $2\zh\chi= Q^2$ have no classical limit, which is why we do not discuss them in the following.\label{footnoteCond}} Among the remaining parameters, we consider $(\tilxi{n})_{n\in\mathbb{N}}$ and $(\tiltheta{n})_{n\in\mathbb{N}}$ along with $\zh$, $Q$ and $\chi$ as (physical) input parameters to the black hole geometry (outside of the horizon) and thus the EMD coefficients. Indeed, the coefficients $(\widetilde{a}_n)_{n\in\mathbb{N}}$ (and thus through (\ref{SeriesInversion}) also $(\widetilde{b}_n)_{n\in\mathbb{N}}$) are fixed by expanding (\ref{FreeFallingTime}) order by order in $(z-\zh)$. Notably, using (\ref{Condxi0}), we find to leading order $\mathcal{O}((z-\zh)^1)$ 
\begin{align}
\widetilde{a}_1=\sqrt{\frac{\tilxi{1}(Q^2-2\zh\chi)^2+2\widetilde{a}_1\,\zh(Q^2-\zh\chi)}{\tiltheta{1}(Q^2-2\zh\chi)^2+2\widetilde{a}_1\,\zh(Q^2-\zh\chi)}}\,,\label{DiffEq1}
\end{align}
where we have implicitly assumed that the right hand side is well-defined, which, as we shall see in the following Subsection, is related to the behaviour of the subleading terms in (\ref{fhzExpansion}): concretely, it is related to the non-vanishing of the terms of order $\mathcal{O}(z-\zh)$ and thus assumes that $z=\zh$ constitutes a \emph{simple} horizon. In this case, $\widetilde{a}_1$ is fixed in terms of $(\tilxi{1}\,,\tiltheta{1}\,,Q\,,\zh\,,\chi)$.\footnote{Although (\ref{DiffEq1}) has multiple solutions, generally only one of them leads to $\widetilde{a}_1\in\mathbb{R}$.} Higher orders $\mathcal{O}((z-\zh)^p)$ (for $p>1$) in (\ref{FreeFallingTime}) only depend on $\widetilde{a}_{p'}$ with $p'\leq p$ and are linear in $\widetilde{a}_p$. These equations therefore uniquely fix the expansion (\ref{TauExpansion}), and therefore the black hole geometry (see also \cite{SHinprogress} for a more detailed discussion in a related case). The proper time allows therefore for an EMD that is capable of describing the charged black hole geometry. We also remark in passing that the condition for $\widetilde{a}_1\in\mathbb{R}$ in (\ref{DiffEq1}) as well $f(z)>0$ and $h(z)>0$ for $z>\zh$ requires
\begin{align}
&(Q^2-2\zh\chi)\,\tilxi{1}>-2\widetilde{a}_1\,\zh\,\frac{Q^2-\zh\chi}{Q^2-2\zh\chi}\,,&&\text{and} &&(Q^2-2\zh\chi)\,\tiltheta{1}>-2\widetilde{a}_1\,\zh\,\frac{Q^2-\zh\chi}{Q^2-2\zh\chi}\,,
%\text{sign}\left(\tilxi{1}(Q^2-2\zh\chi)+2\widetilde{a}_1\,\zh(Q^2-\zh\chi)\right)=\text{sign}\left(\tiltheta{1}(Q^2-2\zh\chi)+2\widetilde{a}_1\,\zh(Q^2-\zh\chi)\right)=\text{sign}(Q^2-2\zh\chi)\,.
\end{align}
with $\widetilde{a}_1$ fixed from (\ref{DiffEq1}). Finally, unlike the EMD based on the spatial distance, finiteness of curvature invariants such as the Ricci- and Kretschmann scalar pose no further constraints on the parameter space, since the second derivatives of $f$ and $h$ are finite at the horizon.

%%%%%%%%%%%%
\subsection{Extremal Black Holes}
\subsubsection{General Charged Extremal Black Holes}\label{Sect:GeneralCharge}
In the previous Subsection, we have argued for an EMD of charged, spherically symmetric black holes in terms of the proper time (\ref{FreeFallingTime}): indeed, by inserting \eqref{zExpansiontau} and \eqref{TauExpansion}, eq.~\eqref{FreeFallingTime} can be expanded in powers of $(z-\zh)$, which provides conditions on the coefficients $(\widetilde{a}_n)_{n\in\mathbb{N}}$ and in fact fixes them iteratively in terms of $\left((\tilxi{n})_{n\in\mathbb{N}}\,,(\tiltheta{n})_{n\in\mathbb{N}}\,,Q\,,\zh\,,\chi\right)$. Here we are interested in the question, which regions of this parameter space allow for \emph{extremal} (charged) black holes. Since these are characterised by the fact that $z=\zh$ is a double zero of $f$ and $h$ in (\ref{fhFunctions}), we have to impose in addition to (\ref{Condxi0}) also the vanishing of the term of order $\mathcal{O}(z-\zh)$ in (\ref{fhzExpansion}): 
\begin{align}
&\frac{\tilxi{1}(Q^2-2\zh\chi)^2+2\widetilde{a}_1\,\zh(Q^2-\zh\chi)}{\widetilde{a}_1\,\zh^2 (Q^2-2\zh\chi)}=0\,,&&\text{and} &&\frac{\tiltheta{1}(Q^2-2\zh\chi)^2+2\widetilde{a}_1\,\zh(Q^2-\zh\chi)}{\widetilde{a}_1\,\zh^2 (Q^2-2\zh\chi)}=0\,.\label{ExtremConstraint}
\end{align}
Imposing these conditions renders the right hand side of (\ref{DiffEq1}) ill-defined and indeed, the leading order term $\mathcal{O}((z-\zh)^1)$ of (\ref{FreeFallingTime}) instead becomes
\begin{align}
\widetilde{a}_1=\sqrt{\frac{\widetilde{a}_1^2\, Q^4+(Q^2-2\zh\chi)\left(\tilxi{2}(Q^2-2\zh\chi)^2+2\widetilde{a}_2\zh(Q^2-\zh\chi)\right)}{\widetilde{a}_1^2\, Q^4+(Q^2-2\zh\chi)\left(\tiltheta{2}(Q^2-2\zh\chi)^2+2\widetilde{a}_2\zh(Q^2-\zh\chi)\right)}}\,.\label{ExtremalDiffEq}
\end{align}
This equation is qualitatively quite different from (\ref{DiffEq1}) and (in conjunction with (\ref{ExtremConstraint})) the type of possible solutions depend more heavily on the remaining parameters. Recalling that we still consider $Q^2\neq 2\zh \chi$, we may distinguish the following subspaces of the parameter space:
\begin{enumerate}
\item[\emph{(i)}]  $Q^2\neq \zh \chi$: In this case, the parameter subspace can further be subdivided by considering (see Appendix \ref{Appx:A} for more details)
\begin{itemize}
\item $\tilxi{2}=\tiltheta{2}$: In this case, the only solution of (\ref{ExtremalDiffEq}) is $\widetilde{a}_1=1$ such that (\ref{ExtremConstraint}) can be used to fix
\begin{align}
\tilxi{1}=\tiltheta{1}=-\frac{2\zh(Q^2-\zh\chi)}{(Q^2-2\zh\chi)^2}\,.
\end{align}
Higher order terms $\mathcal{O}((z-\zh)^p)$ (for $p>1$) in the expansion of (\ref{FreeFallingTime}) are linear in $\widetilde{a}_p$ (and only contain $\widetilde{a}_{p'}$ with $p'\leq p$), thus uniquely fixing the expansion (\ref{TauExpansion}) and therefore allowing to define an EMD for an extremal charged black hole, which is parametrised by $\left((\tilxi{n})_{n\geq 2}\,,(\tiltheta{n})_{n\geq 2}\,,Q\,,\zh\,,\chi\right)$. 
\item $\tilxi{2}\neq \tiltheta{2}$: In this case, $\widetilde{a}_2$ does not disappear from the equation (\ref{ExtremalDiffEq}). Moreover, all higher order terms $\mathcal{O}((z-\zh)^p)$ (for $p>1$) in the expansion of (\ref{FreeFallingTime}) also acquire a linear dependence on $\widetilde{a}_{p+1}$ (but remain independent of all $\widetilde{a}_{p'}$ with $p'> p+1$).\footnote{We have checked up to $p=5$, that the term $\mathcal{O}((z-\zh)^p)$ in (\ref{FreeFallingTime}) depends linearly on $\widetilde{a}_{p+1}$, provided that $\tilxi{2}\neq \tiltheta{2}$ and $Q^2\neq \zh \chi$ (as well as $Q^2\neq 2\zh \chi$).} We therefore use the condition (\ref{ExtremConstraint}) to fix
\begin{align}
&\tiltheta{1}=\tilxi{1}\,,&&\text{and} &&\widetilde{a}_1=\tilxi{1}\left(2\chi-\frac{Q^4}{2\zh(Q^2-\zh\chi)}\right)\,,
\end{align}
leaving $\tilxi{1}$ as an undetermined free parameter. Eq.~(\ref{ExtremalDiffEq}) then determines $\widetilde{a}_2$ and the higher order terms $\mathcal{O}((z-\zh)^p)$ in (\ref{FreeFallingTime}) uniquely fix $\widetilde{a}_{p+1}$ uniquely, without further constraints on any of the $\tilxi{n}$ and $\tiltheta{n}$, thus leading to an EMD of an extremal charged black hole parametrised by $\left((\tilxi{n})_{n\geq 1}\,,(\tiltheta{n})_{n\geq 2}\,,Q\,,\zh\,,\chi\right)$. 
\end{itemize}
\item[\emph{(ii)}] $Q^2= \zh \chi$: In this case, (\ref{ExtremConstraint}) is solved by $\tilxi{1}=0=\tiltheta{1}$, while (\ref{ExtremalDiffEq}) fixes $\widetilde{a}_1$ (irrespective of $\tilxi{2}=\tiltheta{2}$ or not). Assuming the latter to be non-negative, there are two possible solutions
\begin{align}
\widetilde{a}_1=\frac{1}{\sqrt{2}}\,\sqrt{1+\zh\chi\,\tiltheta{2}\pm\sqrt{(1+\zh \chi\tiltheta{2})^2-4\zh\chi\tilxi{2}}}\,.\label{Sola1GnDef}
\end{align}
For $\tilxi{2},\tiltheta{2}\to 0$ (\emph{i.e.} in the classical case), these become $\widetilde{a}_1=1$ and $\widetilde{a}_1=0$ respectively. We furthermore assume $\widetilde{a}_1$ to be real, which in particular requires
\begin{align}
&\tilxi{2}\leq \frac{(1+\zh\chi\tiltheta{2})^2}{4\zh\chi}\,,&&\text{and} &&\tiltheta{2}\geq -\frac{1}{\zh\chi} \hspace{0.2cm} \text{if} \hspace{0.2cm} \tilxi{2}>0\,, 
\end{align} 
The higher order terms $\mathcal{O}((z-\zh)^p)$ in (\ref{FreeFallingTime}) provide linear relations to uniquely fix $\widetilde{a}_{p}$, without further constraints on any of the $\tilxi{n}$ and $\tiltheta{n}$. The region in parameter space $Q^2= \zh \chi$ therefore leads to an EMD of an extremal charged black hole parametrised by $\left((\tilxi{n})_{n\geq 2}\,,(\tiltheta{n})_{n\geq 2}\,,\zh\,,\chi\right)$. 
\end{enumerate}
Furthermore, in all cases, the remaining free parameters still need to be chosen such that
\begin{align}
&f(z)>0\,,&&\text{and} &&h(z)>0\,,&&\forall z\geq \zh\,,\label{PositiveSignature}
\end{align}
which gives rise to further non-trivial inequalities. However, finiteness of curvature invariants such as the Ricci scalar pose no further constraints also in the extremal case, since the second derivatives of $f$ and $h$ remain finite at the horizon.
%%%%%%%%%%
\subsubsection{Deformed Extremal Reissner-Nordstr\"om Geometry}\label{DeformedComputations}
Since the above conditions are complicated to represent in full generality (see Appendix~\ref{Appx:A}), we shall here analyse them in the case of a BH that represents a small deformation of the extremal Reissner-Nordstr\"om geometry. In order to parametrise this case within the remaining parameter space, we introduce
\begin{align}
&\epsilon:=\frac{\chi-\zh}{\zh}\,,&&\tilxi{n}:=\frac{x_n}{\chi^n}\,,&&\tiltheta{n}:=\frac{t_n}{\chi^n}\,,\label{DefSmall}
\end{align}
where $\epsilon>-1$ is understood as a small deformation parameter. Notice in particular that $\zh=\frac{\chi}{1+\epsilon}$. In the following, we shall discuss the case \emph{(i)} to leading order in $\epsilon$, while \emph{(ii)} can be compactly discussed to all orders for $\epsilon>-1$:
\begin{enumerate}
\item[\emph{(i)}] In order for a black hole in the parameter space with $Q^2\neq \zh\chi$ to represent a small deformation of the extremal RN geometry, we need to consider $\epsilon$ to be a small parameter and shall discuss effects to leading order. In particular, we shall assume
\begin{align}
&x_n=\epsilon\,x_n^{(1)}+\mathcal{O}(\epsilon^2)\,,&&\text{and} &&t_n=\epsilon\,t_n^{(1)}+\mathcal{O}(\epsilon^2)\,,\hspace{0.2cm}\forall n\geq 1\,,&&|Q|=\chi(1+q\epsilon)+\mathcal{O}(\epsilon^2)\,,\label{DeformEps}
\end{align}
which allows to write for the coefficients in (\ref{TauExpansion})
\begin{align}
&\widetilde{a}_n=\widetilde{a}_n^{(0)}+\epsilon\,\widetilde{a}_n^{(1)}+\mathcal{O}(\epsilon^2)\,,&&\text{with} &&\left\{\begin{array}{l}\widetilde{a}_1^{(0)} = 1\,, \\ \widetilde{a}_2^{(0)}=0\,, \\ \widetilde{a}_3^{(0)}=-\frac{1}{6\chi^2}\,. %\\ \widetilde{a}_4^{(0)}=-\frac{1}{4\chi^3}\,.  
\end{array}\right.
\end{align}
Furthermore, to leading order in $\epsilon$, eq.~(\ref{Condxi0}) and (\ref{ExtremConstraint}) impose respectively
\begin{align}
&\tilxi{0}=\tiltheta{0}=1+2q\,\epsilon+\mathcal{O}(\epsilon^2)\,,&&\text{and} &&t_1^{(1)}=-2(1+2q)=x_1^{(1)}\,,
\end{align}
while (\ref{ExtremalDiffEq}) (and higher orders in the expansion of \eqref{FreeFallingTime}) uniquely fix the leading deformations $\widetilde{a}_n^{(1)}$, \emph{i.e.} for all values of $x_2^{(1)}$ and $t_2^{(1)}$ we find at leading order in $\epsilon$
\begin{align}
&\widetilde{a}_1^{(1)}=\frac{t_2^{(1)}-x_2^{(1)}}{2}\,,\hspace{2cm}\widetilde{a}_2^{(1)}=\frac{2(t_2^{(1)}-x_2^{(1)})+t_3^{(1)}-x_3^{(1)}}{4\chi}\,,\nonumber\\
&\widetilde{a}_3^{(1)}=-\frac{36+48q+t_2^{(1)}-7x_2^{(1)}-12(t_3^{(1)}-x_3^{(1)})-6(t_4^{(1)}-x_4^{(1)})}{36\chi^2}\,.
\end{align}
Finally, the conditions (\ref{PositiveSignature}) become to leading order
\begin{align}
&1+(6+8q-x_2^{(1)})\epsilon>0\,,&&\text{and} &&1+(6+8q-t_2^{(1)})\epsilon>0\,.\label{CondLeading}
\end{align}

\item[\emph{(ii)}] For the RN geometry, the extremality condition corresponds to $Q^2=\chi^2$, which leads to $\zh=\chi$ ($\epsilon = 0$) and thus $Q^2= \zh \chi$. The latter relation can therefore be understood as a deformation of the classical case for any value of $\epsilon>-1$. Furthermore, since $\widetilde{a}_1=1$ classically, we shall focus on the positive sign in~(\ref{Sola1GnDef}): 
\begin{align}
\widetilde{a}_1=\frac{\sqrt{1+\epsilon+t_2+\sqrt{(1+\epsilon+t_2)^2-4(1+\epsilon)x_2}}}{\sqrt{2(1+\epsilon)}}
\end{align}
Finally, the conditions (\ref{PositiveSignature}) can be formulated as follows
\begin{align}
&\left\{\begin{array}{lcl}x_2<0 & \text{if} & t_2\leq -(1+\epsilon)\,, \\ x_2\leq \frac{(1+\epsilon+t_2)^2}{4(1+\epsilon)} & \text{if} & -(1+\epsilon)<t_2<1+\epsilon\,, \\ x_2<t_2 & \text{if} & t_2\geq 1+\epsilon\,.\end{array}\right.
%
%&\left\{\begin{array}{lcl}x_2<0\leq \frac{(1+\epsilon+t_2)^2}{4(1+\epsilon)} & \text{if} & -(1+\epsilon)<t_2\leq 0\,, \\ t_2<x_2\leq \frac{(1+\epsilon+t_2)^2}{4(1+\epsilon)} & \text{if} & 0<t_2<1+\epsilon\,.\end{array}\right.&&\widetilde{a}_1=\frac{\sqrt{1+\epsilon+t_2-\sqrt{(1+\epsilon+t_2)^2-4(1+\epsilon)x_2}}}{\sqrt{2(1+\epsilon)}}\,.
\end{align}
These conditions are represented by the green region in Figure~\ref{Fig:TypeiiExtremal}: the interior satisfies the inequalities (\ref{PositiveSignature}), while only the solid black line of its boundary is compatible. 
\begin{figure}[h]
\begin{center}
\includegraphics[width=8cm]{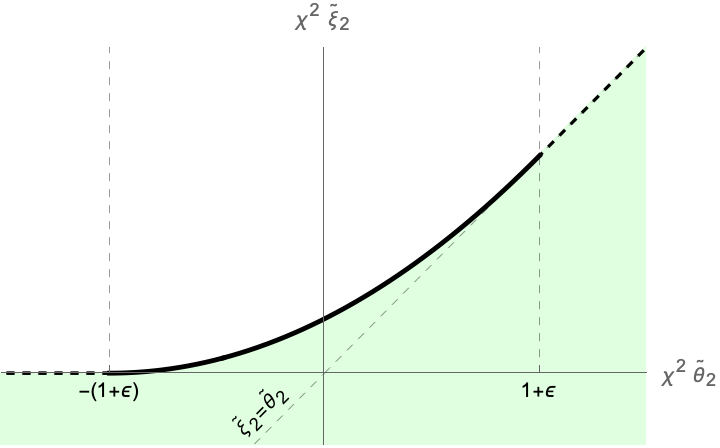} 
\end{center}
\caption{Allowed parameter region of $\tilxi{2}$ and $\tiltheta{2}$ in the case $Q^2=\zh\,\chi$. The interior of the green coloured region leads to extremal geometries satisfying all relevant conditions. On its boundary, only the solid black line is allowed, while the dashed black line is excluded. }
\label{Fig:TypeiiExtremal}
\end{figure}

\end{enumerate}

\subsection{Applications}
Before closing this Section, we present two applications of our framework: first, as a concrete example, we shall discuss the Frolov space-time \cite{Frolov:2016pav} in the extremal as a deformation of the Reissner-Nordstr\"om geometry. Secondly, we discuss a connection of our framework to the  Weak Gravity Conjecture within the Swampland program \cite{Palti:2019pca,Vafa:2005ui,Arkani-Hamed:2006emk}.

\subsubsection{Extremal Frolov space-time}
As a concrete example of the EMD based on the proper time (notably in the extremal case), we shall consider here the following metric (using the same notation as in \eqref{MetricForm}), first introduced by Frolov in~\cite{Frolov:2016pav}
%\begin{align}
%\label{FrolovMetric}
%    \dd s^2 = - F(z) \dd t^2 + \frac{\dd z^2}{F(z)} + z^2 \dd \theta^2 + z^2 \sin^2 \theta \dd \phi^2 \ , \quad F(z)= 1 - \frac{(2 \chi z - Q^2) z^2}{z^4 + (2 \chi z + Q^2) \eta^2} \ .
%\end{align}
\begin{align}
\label{FrolovMetric}
    h(z)=f(z)= 1 - \left(\frac{2\chi}{z}-\frac{Q^2}{z^2} \right)\widetilde{\Gamma} \ , \qq{where} \widetilde{\Gamma}\coloneqq\frac{z^4}{z^4 + (2 \chi z + Q^2) \eta^2} \ .
\end{align}
If $Q=0$, we recover the Hayward metric \cite{Hayward_2006}. Inspecting the two limits
\begin{align}
    f(z) &\overset{z\to \infty}{=} 1 - \frac{2 \chi}{z} + \frac{Q^2}{z^2} + \eta^2 \mathcal{O}(z^{-4}) &\text{and} & & f(z) &\overset{z\to 0}{=} 1 + \frac{z^2}{\eta^2} + \mathcal{O}(z^6) \ .
\end{align}
the metric (\ref{FrolovMetric}) features asymptotic flatness, while reproducing the Reissner-Nordström metric for $\eta\to 0$. This latter parameter can be interpreted as the radius of an Anti de-Sitter core at the origin. Finally, the position of the event horizon satisfies $f(\zh)=0$, which has 4 solutions: we shall understand $\zh$ to be the largest (real) value among these.
%\begin{align}
%\zh=\chi\left(1+\sqrt{1-Q^2/\chi^2}\right)+\frac{\eta^2}{2Q^2}\left(4\chi+\frac{Q^2-4\chi^2}{\chi\sqrt{1-Q^2/\chi^2}}\right)+\mathcal{O}(\eta^4)\,.
%\end{align}

\begin{figure}[h]
\begin{center}
\includegraphics[width=7.5cm]{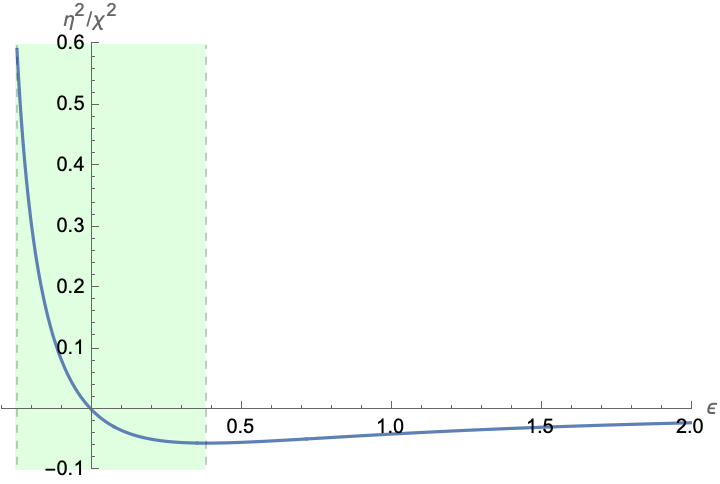} \hspace{0.5cm}\includegraphics[width=7.5cm]{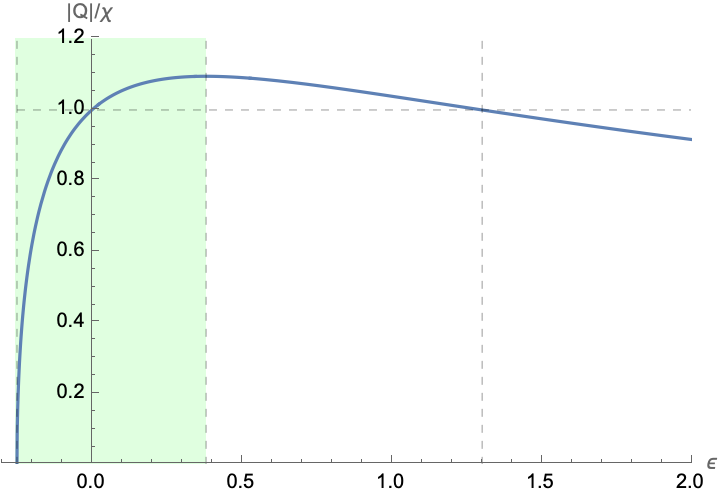}
\end{center}
\caption{Parametrisation of the radius of the Anti de-Sitter core (left) and mass-to-charge ratio (right) of the extremal Frolov-space-time as a function of the deformation parameter $\epsilon$ (see eq.~(\ref{FormFrolovQeta})). The green shaded region leads to a geometry with the correct signature outside of the event horizon. Notice, however, that $\epsilon>0$ leads to $|Q|>\chi$ and $\eta^2<0$, such that the latter is no longer the radius of a Anti de-Sitter core.}
\label{Fig:ExtremalFrolov}
\end{figure}

In the following, we shall be interested in the extremal Frolov space-time. Indeed, extremality ({\emph{i.e.}} $\frac{df}{dz}(\zh)=0$) imposes a condition among the three parameters $(Q,\chi,\eta)$. In order to characterise this condition, we shall consider $\eta$ to be parametrically small, by introducing the parameter $\epsilon$ as in (\ref{DefSmall}) to capture small deformations of the Reissner-Nordstr\"om geometry
\begin{align}
&\epsilon:=\frac{\chi-\zh}{\zh}&&\text{such that}&&\zh=\frac{\chi}{1+\epsilon}\,,
\end{align}
where $\zh=\chi$ corresponds to the classical event horizon of the (extremal) Reissner-Nordstr\"om geometry. The conditions for a (deformed) extremal black hole are then solved by
\begin{align}
&\frac{|Q|}{\chi}=\frac{\sqrt{\epsilon+\sqrt{1+5\epsilon+5\epsilon^2}}}{1+\epsilon}\,,&&\frac{\eta^2}{\chi^2}=\frac{1+2\epsilon-\sqrt{1+5\epsilon+5\epsilon^2}}{(1+\epsilon)^3}\,.\label{FormFrolovQeta}
\end{align}
Reality of $Q$ furthermore requires $\epsilon\geq -\frac{1}{4}$, while $f>0$ for $z>\zh$ requires $\epsilon\leq\frac{5+\sqrt{105}}{40}$. These results are schematically shown in Figure~\ref{Fig:ExtremalFrolov}, where the green shaded region $\epsilon\in[-1/4,\frac{5+\sqrt{105}}{40}]$ denotes the parameter space that gives rise to an extremal black hole with the correct space-time signature outside of the event horizon. 
We remark that $\eta^2>0$ for $\epsilon\in[-1/4,0]$, which correspond to a horizon radius that is larger than its classical counterpart. Allowing $\eta^2<0$, we find $|Q|\geq\chi$ for $\epsilon\in\left[0,\frac{\sqrt{13}-1}{2}\right]$, thus circumventing the classical bound for the charge of the black hole. We remark, however, that $\eta^2<0$, no longer represents the radius of an Anti de-Sitter core at the origin, which, however, is a priori not visible only from outside of the event horizon. Notice, also that $\epsilon=\frac{5+\sqrt{105}}{40}$ corresponds to the largest possible value of $|Q|/\chi$ and the smallest possible value of $\eta^2/\chi^2$ (\emph{i.e.} the extrema of the expressions in (\ref{FormFrolovQeta})). Furthermore we remark that $Q^2=2\chi\zh$ for $\epsilon=3/4>\frac{5+\sqrt{105}}{40}$, which is outside of the green shaded region in Figure~\ref{Fig:ExtremalFrolov}, thus justifying the assumption made in Section~\ref{Sect:GeneralCharge}.

\begin{figure}[h]
\begin{center}
\includegraphics[width=7.5cm]{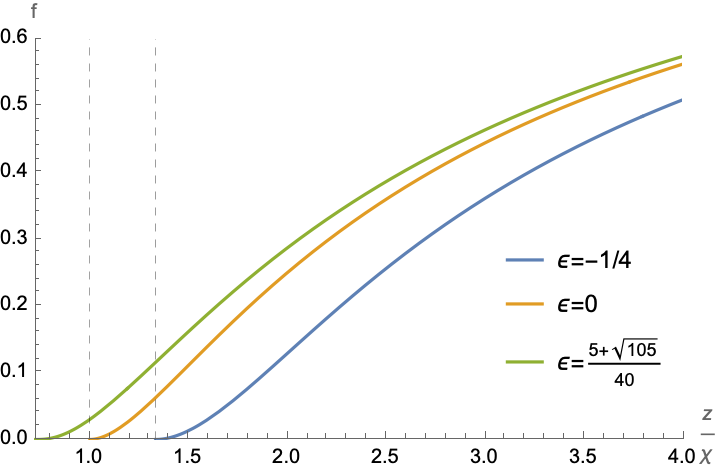} \hspace{0.5cm}\includegraphics[width=7.5cm]{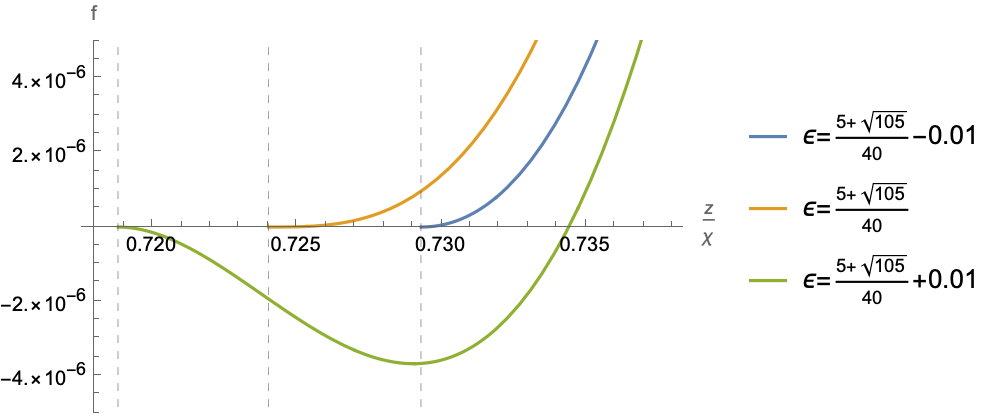}
\end{center}
\caption{Left: plot of the function $f$ outside of the horizon for different values of $\epsilon\in\left[-\tfrac{1}{4},\tfrac{5+\sqrt{105}}{40}\right]$ (\emph{i.e.} the green shaded region in Figure~\ref{Fig:ExtremalFrolov}). Notice that indeed each plot starts with a horizontal tangent, indicating the second order zero at the horizon. Right: for $\epsilon>\frac{5+\sqrt{105}}{40}$, $\zh=\frac{\chi}{1+\epsilon}$ no longer represents the outermost horizon of the geometry.}
\label{Fig:PlotsF}
\end{figure}

Figure~\ref{Fig:PlotsF} depicts the function $f$ for different values of $\epsilon$ outside of the 
event horizon. The right panel of this Figure indicates that indeed for $\epsilon>\frac{5+\sqrt{105}}{40}$, the function $f$ is negative outside of the double horizon located at $\frac{\chi}{1+\epsilon}$: moreover, it develops a further single horizon at a position $z>\chi$.

\begin{figure}[h]
\begin{center}
\includegraphics[width=12cm]{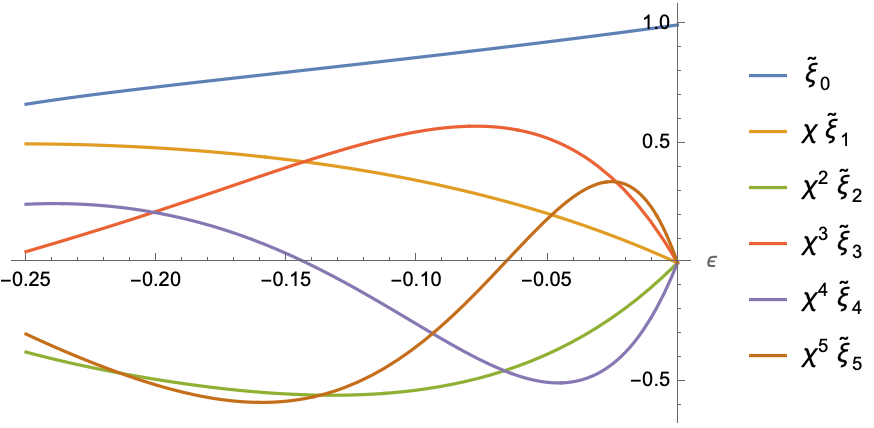}
\end{center}
\caption{Leading Coefficients $\tilxi{n}(\epsilon)$ for $\epsilon\in\left[-\tfrac{1}{4},0\right]$.}
\label{Fig:PlotsXiCoefs}
\end{figure}

Finally, we also provide the expressions for the first few coefficients $(\tilxi{n}=\tiltheta{n})_{n\in\mathbb{N}}$ as functions of $\epsilon$
\begin{align}
&\tilxi{0}=\frac{1}{2+\epsilon -\sqrt{1+5 \epsilon  (1+\epsilon )}}=1+\frac{3\epsilon}{2}+\mathcal{O}(\epsilon^2)\,,\nonumber\\
&\tilxi{1}=\frac{2(1+\epsilon)(1-\sqrt{1+5\epsilon(1+\epsilon)})}{(2+\epsilon-\sqrt{1+5\epsilon(1+\epsilon)})^2\,\chi}=-\frac{5\epsilon}{\chi}+\mathcal{O}(\epsilon^2)\,,\nonumber\\
&\tilxi{2}=\,\frac{2(1+\epsilon)^2(1+10\epsilon+8\epsilon^2-(1+2\epsilon)\sqrt{1+5\epsilon(1+\epsilon))})}{(2+\epsilon-\sqrt{1+5\epsilon(1+\epsilon)})^3\,\chi^2}=\frac{11\epsilon}{\chi^2}+\mathcal{O}(\epsilon^2)\,.
\end{align}
which are plotted (for $\epsilon\in[-1/4,0]$) in Figure~\ref{Fig:PlotsXiCoefs}.

In order to compare to the leading order computations in Section~\ref{DeformedComputations} (\emph{i.e.} for the case \emph{(i)} of the parameter space), we need to expand (\ref{FormFrolovQeta}) to obtain $\frac{|Q|}{\chi}=1+\frac{3\epsilon}{4}+\mathcal{O}(\epsilon^2)$, such that $q=\frac{3}{4}$ in (\ref{DeformEps}). For small enough values of $\epsilon$ (such that the higher orders in $\epsilon$ can indeed be neglected), this is compatible with (\ref{CondLeading}).

\subsubsection{Weak Gravity Conjecture}
To gain insights on $\tilxi{0}$, we consider the decay of an electrically charged black hole.
The weak gravity conjecture \cite{Palti:2019pca, Harlow:2022ich} states that the emission spectrum requires a particle with higher charge-to-mass ratio with respect to black hole itself. In particular, in Planck units, the following relation holds:
\begin{align}
\label{CtMRatio}
    \frac{\chi}{q} \geq \left. 
    \left(\frac{\mathfrak{m}}{\mathfrak{q}} \right) \right|_\text{min} \ , 
\end{align}
where $\mathfrak{m}$ and $\mathfrak{q}$ are the mass and charge of the particle with the largest charge-to-mass ratio of the emission spectrum. In our framework, from the definition of horizon, the ratio between the black hole mass $\chi$ and its charge $q$ is
\begin{align}
    \frac{\chi}{q} = \frac{z_H^2 + q^2 \tilxi{0}}{2  q z_H \tilxi{0}} \ .
\end{align}
Close to the classical extremality condition, $\zh$ and the deformation function $\tilxi{0}$ take the following form:
\begin{align}
\label{NearExtremality}
    z_H = q (1+\gamma) \ , \quad \tilxi{0} =1 + \lambda \ ,
\end{align}
where $\lambda$ and $\gamma$ play the role of the quantum corrections, and $\gamma \sim \order{\lambda}$. 
Using equation \eqref{NearExtremality}, we can obtain the following upper limit:
\begin{align}
    \lambda \leq 2 - 2 \left. \left(\frac{\mathfrak{m}}{\mathfrak{q}}\right) \right| _{\text{min}} \ ,
\end{align}
which relates the inside of the black hole with its emission spectrum, and may allow to grasp some information of the interior structure of the black hole.

%%%%%%%%%%%%%%%%%%%%%%%%%%%%%%%%%%%%

\section*{Conclusions}
In this paper, we studied static and spherically symmetric black hole metrics featuring electric and magnetic charges. Following the approach introduced in \cite{DelPiano:2023fiw}, we extended and further developed the effective metric framework and derived general constraints on its metric coefficients at the horizon. The EMD was built to respect the classical Reissner-Nordstr\"om space-time symmetries and its asymptotics. It also allowed us to determine the general expression for the Hawking temperature of a deformed charged space-time. To gain further insight, we expanded the modified Hawking temperature in power of the charge, and studied the leading order.  We then investigated how the metric deformations impact the electro-magnetic fields. Of special interest is the extremal black hole limit. For this case, we showed that the proper time for an in-falling observer is a better suited physical quantity for the formulation of the EMD.   We test our framework via the Frolov space-time \cite{Frolov:2016pav} and, lastly, we provide a connection between the effective metric coefficients and the Weak Gravity conjecture \cite{Arkani-Hamed:2006emk}.
\\
The formulation of charged black holes within the EMD framework allows to describe phenomena combining the gravitational and electromagnetic interactions in a model independent way. We therefore expect our results to be relevant for future studies, for example for the emission of electromagnetic radiation \cite{Brito:2024tqd} and the stability of charged particle orbits \cite{Baker:2023gdc,Lei:2021koj}, and superradiance effects \cite{Alvarez-Dominguez:2024ahv,Balakumar:2020gli}. The study of these phenomena in a model-independent fashion will give further insights into the interaction of black holes and charged matter, including (quantum) effects beyond General Relativity.

From a conceptual perspective, by including extremal black holes, we have extended the class of black hole geometries that can be described by EMDs. This demonstrates the versatility of our approach, which we hope to extend even further in future work: notably, it would be interesting to include rotational black holes, which would enable us to describe deformations of the Kerr-geometry in an effective fashion. We expect that such a description will have further phenomenological applications~\cite{Maselli:2023khq,EventHorizonTelescope:2019dse}.

\section*{Acknowledgements}
We thank Mikolaj Myszkowski and Vania Vellucci for helpful discussions. S.H. and M. Del Piano would like to thank the Quantum Theory Center (QTC) at the Danish Institute for Advanced Study and IMADA of the University of Southern Denmark for the hospitality. The work of F.S. is partially supported by the Carlsberg Foundation, grant CF22-0922.

%\newpage

\appendix

\begin{comment}
     \section{Coefficient $\widetilde{a}_2$ in the extremal case}\label{Appx:A}
For the completeness of the discussion, we report the expression of the coefficient $\widetilde{a}_2$, which relates among themselves higher derivatives of the metric functions $f$ and $g$, as showed in \eqref{ZinTfallExp}.
\vspace{0.5cm}

\begin{small}
    $\widetilde{a}_2 =\frac{\left(Q^2-2 \chi  z_H\right){}^2 \left(-16 \widetilde{\xi} _2 z_H^4 \left(Q^2-\chi  z_H\right){}^4+\widetilde{\xi} _1^2 \left(Q^2-2 \chi  z_H\right){}^4 \left(4 \widetilde{\theta} _2 z_H^2 \left(Q^2-\chi  z_H\right){}^2+\widetilde{\xi} _1^2 Q^4 \left(Q^2-2 \chi  z_H\right)\right)-4 \widetilde{\xi} _1^2 Q^4 z_H^2 \left(Q^2-\chi  z_H\right){}^2 \left(Q^2-2 \chi  z_H\right)\right)}{8 z_H^3 \left(Q^2-\chi  z_H\right){}^3 \left(4 z_H^2 \left(Q^2-\chi  z_H\right){}^2-\widetilde{\xi} _1^2 \left(Q^2-2 \chi  z_H\right){}^4\right)}$
    \end{small}
\end{comment}

 \section{Conditions in the extremal case with $Q^2 \neq z_H \chi$}\label{Appx:A}
Here, we collect all conditions on the deformation functions coeffcients for an extremal black hole with $\widetilde{\Gamma} \neq \widetilde{\Omega}$ and $Q^2 \neq z_H \chi$ discussed in Subsection \ref{Sect:GeneralCharge}. 
In particular, requiring $\widetilde{a}_1 \in \mathbb{R}$ and \eqref{PositiveSignature}, we have:
\begin{tcolorbox}[ams equation,colback=black!10!white,colframe=black!95!green]
\parbox{15cm}{${}$\\[-40pt] \\ 
\begin{itemize}
    \item[$ \widetilde{\xi}_0$] $ = \widetilde{\theta}_0 = \frac{z_H^2}{2 \chi z_H - Q^2} \ , \quad \widetilde{\xi}_1 = \widetilde{\theta}_1 = \widetilde{a}_1 \frac{2 \widetilde{\xi}_0^2 \left(\chi  z_H-Q^2\right)}{ z_H^3}  \ , \quad \widetilde{\xi}_2 > \widetilde{\theta} _2 \ ,$
    \item  $\frac{Q^2}{2}<  \chi z_H < Q^2 \ , \ \widetilde{a}_2<\frac{z_H \left(4 \widetilde{\theta}_2 z_H^2 \left(Q^2-\chi  z_H\right)^2+\widetilde{\xi}_1^2 Q^4 \left( Q^2 - 2 \chi z_H \right)\right)}{4 \widetilde{\xi}_0^2 \left(\chi  z_H-Q^2\right)^3} \ ,$
    \item $ \chi  z_H<\frac{Q^2}{2} \ , \ \widetilde{a}_2>\frac{z_H \left(4 \widetilde{\xi}_2 z_H^2 \left(Q^2-\chi  z_H\right)^2+\widetilde{\xi} _1^2 Q^4 \left( Q^2 - 2 \chi z_H \right)\right)}{4 \widetilde{\xi}_0^2 \left(\chi  z_H-Q^2\right)^3}  \ , \nonumber$
    \item $\chi  z_H>Q^2 \ , \ \widetilde{a}_2>\frac{z_H \left(4 \widetilde{\theta} _2 z_H^2 \left(Q^2-\chi  z_H\right)^2+\widetilde{\xi} _1^2 Q^4 \left( Q^2 - 2 \chi z_H \right)\right)}{4 \widetilde{\xi}_0^2 \left(\chi  z_H-Q^2\right)^3} \ . \nonumber$
\end{itemize}
 ${}$\\[-30pt]}
\end{tcolorbox}
% For $f^{(2)}_H=h^{(2)}_H$
% \begin{tcolorbox}[ams equation,colback=black!10!white,colframe=black!95!green]
% \parbox{15cm}{${}$\\[-40pt]
% \begin{align}
%     &\widetilde{a}_2 <\frac{\left(Q^2-2 \chi  z_H\right)^2 \left(4 \xi _2 z_H^2 \left(Q^2-\chi  z_H\right)^2+\xi _1^2 \left(Q^6-2 Q^4 \chi  z_H\right)\right)}{4 z_H^3 \left(\chi  z_H-Q^2\right)^3} \qq{and} \frac{Q^2}{2 z_H}<\chi <\frac{Q^2}{z_H} \ , \nonumber \\ \qq{or}&\widetilde{a}_2>\frac{\left(Q^2-2 \chi  z_H\right)^2 \left(4 \xi _2 z_H^2 \left(Q^2-\chi  z_H\right)^2+\xi _1^2 \left(Q^6-2 Q^4 \chi  z_H\right)\right)}{4 z_H^3 \left(\chi  z_H-Q^2\right)^3}\ \qq{and} \nonumber \\ 
%     &\left(2 \chi  z_H<Q^2 \qq{or} \chi  z_H>Q^2\right)
%     \end{align}\nonumber ${}$\\[-30pt]}
% \end{tcolorbox}
\noindent
If $\frac{d^2f}{dz^2}\big|_{z=\zh} = \frac{d^2h}{dz^2}\big|_{z=\zh}$, then $\widetilde{a}_1=1$, and the relation $\widetilde{\xi}_2>\widetilde{\theta} _2$ becomes $\widetilde{\xi}_2 = \widetilde{\theta} _2$, such that:
\begin{tcolorbox}[ams equation,colback=black!10!white,colframe=black!95!green]
\parbox{15cm}{${}$\\[-40pt]
\\
\begin{itemize}
    \item[$\widetilde{\xi}_0$] $ = \widetilde{\theta}_0 = \frac{z_H^2}{2 \chi z_H - Q^2} \ , \quad \widetilde{\xi}_1 = \widetilde{\theta}_1 = \frac{2 \widetilde{\xi}_0^2 \left(\chi  z_H-Q^2\right)}{ z_H^3}  \ , \quad \widetilde{\xi}_2 = \widetilde{\theta} _2 \ , $
    \item $\frac{Q^2}{2 }<\chi z_H \ , \quad \widetilde{\xi} _2<-\frac{Q^4}{\left(Q^2-2 \chi  z_H \right)^3}  , \nonumber$
    \item $  \chi  z_H<\frac{Q^2}{2} \ ,\quad \widetilde{\xi} _2>-\frac{Q^4}{\left(Q^2-2 \chi  z_H \right)^3}  \  . \nonumber$
         \end{itemize}
     ${}$\\[-30pt]}
\end{tcolorbox}

\printbibliography

\end{document}